\documentstyle[pra,aps,epsf]{revtex}
\draft

\begin{document}

\title{Searches for violation of fundamental time reversal and
space reflection symmetries in solid state experiments}

\author{S. A. Kuenzi, O. P. Sushkov, V. A. Dzuba, and J. M. Cadogan}

\address{School of Physics, University of New South Wales,\\
 Sydney 2052, Australia}

\maketitle

\begin{abstract}
The electric dipole moment (EDM) of a particle violates both time reversal
(T) and space reflection (P) symmetries.
There have been  recent suggestions for searches of the electron EDM using
solid state experiments \cite{Lam,Hun}. These experiments could
improve the sensitivity compared to present atomic and molecular experiments 
by several orders of magnitude.
In the present paper we calculate the expected effect.
We also suggest that this kind of experiment is sensitive to T,P-violation
in nuclear forces and calculate effects caused by the nuclear Schiff
moment.

The compounds under consideration contain magnetic Gd$^{3+}$ ions and
oxygen O$^{2-}$ ions. We demonstrate that the main mechanism for the
T,P-odd effects is related to the penetration of the Oxygen 2p-electrons to 
the Gd core. All the effects are related to the deformation of the 
crystal lattice.

\end{abstract}


\section{introduction}
Violation of the combined symmetry of charge conjugation (C) and parity (P)
has been discovered in the decay of the $K^0$ meson about 40 years ago 
\cite{CCK}. The exact origin of this symmetry violation remains an enigma, 
although the so called standard model of electroweak interactions can 
describe these processes phenomenologically. It has also been proposed by 
Sakharov \cite{Sak} that the matter-antimatter asymmetry observed in our 
Universe could have arisen from a CP-violating interaction active at an 
early stage of the Big Bang. The CP-violation implies a time-reversal (T) 
asymmetry and hence violation of both T- and P-symmetries, because there are 
strong reasons to  believe that the combined CPT-symmetry should not be 
violated \cite{CPT}.
Electric dipole moment (EDM) of a system in a stationary quantum state 
indicates violation of  T- and P-symmetries. This is why searches for 
EDM of elementary particles, atoms and molecules are very important for
studies of  violations of fundamental symmetries \cite{KL}.
A property that is closely related to EDM is the so called Schiff moment,
see e.g. Ref. \cite{KL}. A non zero Schiff moment also indicates 
T- and P-violation. In the present paper we consider T,P-odd effects in
solids due to the EDM of the electron and due to the nuclear Schiff moment.

The present best limitation on the electron EDM comes from the Berkeley 
experiment with an atomic Thallium beam \cite{Com}, 
\begin{equation}
\label{Tl}
d_e < 1.6\times 10^{-27}e\ cm.
\end{equation}
There are some ideas for improvement of the sensitivity.
One way of improvement is an experiment with metastable levels of PbO 
molecules \cite{DeM}.
A  breakthrough could be achieved in solid state experiments with
compounds containing uncompensated spins.
This idea was suggested by Shapiro in 1968 \cite{Sh}.
Application of a strong electric field to electrons bound within a
solid would align the EDMs of these electrons. This
should lead to a simultaneous alignment of the electron spins; the 
magnetic field arising from this alignment could be detected
experimentally. Another possibility is to polarize electrons by the
external magnetic field. This causes alignment of electron EDMs, and
hence induces a voltage across the sample that could be detected.
An experiment of this kind has been performed with
nickel-zinc ferrite \cite{VK}. However due to experimental
limitations the result was not very impressive. Interest in this approach
has been renewed recently due to a suggestion of Lamoreaux \cite{Lam} and 
Hunter \cite{Hun} to perform similar experiments with Gadolinium Gallium
Garnet, Gd$_3$Ga$_5$O$_{12}$ (GGG), and Gadolinium Iron Garnet 
Gd$_3$Fe$_5$O$_{12}$ (GdIG)
employing new experimental techniques. The estimates of sensitivity look
highly promising. In the present work we calculate the expected effects.

The best limitation on the Nuclear Schiff moment (NSM) comes from the 
Seattle experiment with atomic $^{199}$Hg \cite{Rom}, 
\begin{equation}  \label{Hg}
   S_N(^{199}Hg) < 0.7 \times 10^{-49}e \ cm^3.
\end{equation}
The expected value of $S_N$ expressed in terms of the fundamental
CP-violating interaction depends on the structure of the particular 
nucleus, 
however for heavy nuclei this dependence is not that strong and therefore we 
will take the value (\ref{Hg}) as a reference point. Later, we will 
comment on a possible additional enhancement due to a special structure of the 
$^{155}$Gd nucleus.
One can search for NSM in the  experiments with GGG and GdIG
simultaneously with searches for the electron EDM. The point is  
that due to the hyperfine interaction nuclear polarization causes polarization
of electrons and vice versa. Another possibility is to use a compound without
unpaired electrons, for example Lutetium Gallium Garnet, Lu$_3$Ga$_5$O$_{12}$
or Lutetium Oxide Lu$_2$O$_3$. In this case one can cool the nuclear spin 
subsystem down
to very low temperature \cite{NT} and this gives an additional enhancement of
the effect. In the present work we calculate the coefficient of proportionality
between voltage and nuclear polarization due to NSM.

Structure of the paper is the following. In Section II we present simple
estimates of the effects induced by the electron EDM.
Section III contains  derivation of the effective interaction induced by
the electron EDM. The Gd$^{3+}$ octupole moment is also discussed here.
In Section IV we calculate wave functions of Oxygen electrons penetrating
inside the Gd$^{3+}$ core. Section V presents derivation of the effective
Hamiltonian which relates the electron EDM with deformation of the lattice.
In Sections VI and VII we derive final results for the voltage across the 
sample and the energy shift in the external electric field induced by
the electron EDM. Section VIII presents similar results for effects
induced by the Gd Nuclear Schiff Moment. Finally Sections IX and X
contain estimates of accuracy of the calculations and our conclusions.

\section{Simple estimates for effects related to the electron EDM}
In this section we follow simple and  important estimates performed
in Ref. \cite{Lam}. GdIG is an ionic crystal  consisting of 
Gd$^{3+}$, Fe$^{3+}$ and O$^{2-}$ ions, see Ref. \cite{GGG}. 
Uncompensated electronic spins are 
localized at Gd$^{3+}$ which has a $4f^7$ electronic configuration and  
Fe$^{3+}$ which has a $4d^5$ electronic configuration.
Both these ions have half-filled electronic shells, hence the orbital angular
momenta of these shells are zero. This explains a very weak spin anisotropy
in this compound. If the electron EDM is not zero, then both Gd$^{3+}$ and
Fe$^{3+}$ have  induced EDMs proportional to $d_e$, $d_a=Kd_e$. It is known
that the coefficient $K$ scales as $Z^3$, where Z is the nuclear charge, see, 
e.g. \cite{KL}. Therefore we neglect the EDM of Fe$^{3+}$ and consider only 
Gd$^{3+}$. GGG differs from GdIG  in the replacement of magnetic  Fe$^{3+}$ by
nonmagnetic  Ga$^{3+}$ ions. The most important Gd$^{3+}$ ion is exactly
the same.
The enhancement coefficient for  Gd$^{3+}$ has been calculated previously
\cite{Stefan}
\begin{equation} \label{K}
   d_a=Kd_e, \ \ \ K=-4.9 \pm 1.6
\end{equation}
%
If all the uncompensated spins in the sample are 100\% polarized then the 
induced electric field in the sample is $E=4\pi n_{Gd}d_a$, where 
$n_{Gd}=1.235 \times 10^{22} cm^{-3}$ is the number density of Gd. This gives
the following induced voltage across an $L=10cm$ sample
\begin{equation}  \label{naive}
   V=EL=4\pi n_{Gd} K d_e L = 2 \times 10^{-9}V.
\end{equation}
This numerical estimate corresponds to the current limitation on $d_e$, see
(\ref{Tl}) and would be absolutely correct for a crystal consisting of
neutral atoms or molecules. However the charge of Gd$^{3+}$ is not zero.
Hence the average electric field acting on the ion 
must be equal to zero,
and hence the EDM of the ion cannot really produce the field (\ref{naive}).
This means that the crystal lattice relaxes in such a way that the effect
(\ref{naive}) is canceled out. This is the Schiff theorem \cite{Sch}.
The theorem is violated due to the finite size of the ion. It means
that the estimate (\ref{naive}) should be reduced by some factor depending
on the ratio of the ion size to the lattice spacing. Calculation of this
effect is performed in the following sections.

On can also perform an estimate of the resulting magnetization 
due to an applied electric field $E_0$ to the sample.
The field inside the solid is $E=E_0/\epsilon$ where
$\epsilon$ is the dielectric constant
(for GdIG $\epsilon \approx 15$, and for GGG $\epsilon \approx 30$ see Ref.
\cite{epGIG})
The T,P-odd energy shift per Gd ion is given by,
\begin{equation} \label{n2}
\delta\epsilon =   d_a  E  = Kd_e  E   = 0.9\times 10^{-22}eV.
\end{equation}
The numerical estimate corresponds to the current limitation on $d_e$, see
(\ref{Tl}) and $E=10kV/cm$. Using Boltzmann distribution one can deduce the
resulting magnetization of the medium.
Similar to (\ref{naive}) the estimate (\ref{n2}) contradicts the
Schiff theorem. An accurate calculation of $\delta \epsilon$ is performed 
below.

\section{Interaction of the Gd$^{3+}$ ion with environment,\\
 Vanishing of the octupole moment of the ion}

In this section we follow the method of calculation suggested in Ref. 
\cite{SFK} (see also \cite{KL}). Let 
\begin{equation} \label{f1}
   \phi({\bf r})=\int d^3R\,{{eq(\bf R)}\over{|{\bf r-R}|}}
   -{\bf r \cdot E}_{ext}
\end{equation}
be the electric potential produced by the environment of a particular
Gd$^{3+}$ ion, $q(\bf R)$ is the  charge density of surrounding ions 
and ${\bf E}_{ext}$ is 
the electric field produced by distant ions and external sources.

The energy correction for a system which has a non-zero EDM is
\begin{equation} \label{f2}
   \Delta \epsilon = e \int d^3r \, \delta \rho ( \bf r )
      \phi(\mathbf{r}) .
\end{equation}
where $\delta \rho(\bf r)$ is the correction to the charge 
 density of Gd$^{3+}$ due to the T,P-odd interaction $H_{TP}$. 

In equilibrium the average force acting on Gd$^{3+}$ vanishes,
\begin{equation} \label{f3}
   \langle F_{Gd^{3+}}\rangle
         = - e\int d^3r\; \rho_0 {\bf{\nabla}}_r \phi( {\bf r})
	 =0.
\end{equation}
Here $\rho_0(\bf r )$ is the spherically symmetric charge density of
Gd$^{3+}$ normalized by the condition $\int d^3r \, \rho_0({\bf r})=3$.
The total charge density of Gd$^{3+}$ is given by 
$\rho(\bf r) = \rho_0(\bf r) + \delta \rho( r)$.

Using eqs.~(\ref{f2}) and (\ref{f3}) and assuming that ${\bf E}_{ext}$ is 
uniform in the vicinity of Gd$^{3+}$ one can rewrite the energy correction as
\begin{eqnarray} \label{f4}
   \Delta \epsilon &=& e \int d^3r\,  \delta \rho({\bf r})
   \phi({\bf r})
   - \frac{1}{3} 
     \int {d^3r} \rho_0({\bf r}) ({\bf d_a \cdot \nabla}_r)
\phi({\bf r}) \nonumber\\ 
     &=& e^2 \int d^3R \, q({\bf R})  \int d^3r \; \left(
     \frac{\delta\rho({\bf r})}
     { \left| {\bf r - R} \right|}
  +\frac{1}{3}
({\bf d_a \cdot \nabla}_R)     
            \frac{\rho_0({\bf r})}
       { \left|{\bf{r} - R} \right|}
   \right).
\end{eqnarray}
 
The charge density correction  $\delta \rho(\bf r)$  can be expanded in 
a series of spherical harmonics,
\begin{equation}
\label{ff}
   \delta \rho({\bf r}) 
   = \sum_{lm} f_{lm}(r) Y_{lm}( \theta_r, \phi_r),
\end{equation}
where
\begin{equation} \label{f6}
   f_{lm}(R) \propto \sum_n
   \frac{\langle 0 | \delta(R-r) Y_{lm} | n \rangle 
   \langle n|H_{TP}|0 \rangle}
   {E_0 - E_n} \; .
\end{equation}
Here we denote the argument of $f_{lm}$ by $R$ to distinguish it from the
internal variable $r$ in the matrix elements.
Because of the negative parity of $H_{TP}$ only odd $l$ contribute
to (\ref{ff}). The $Y_{1m}$ terms correspond to the EDM of Gd$^{3+}$, and 
$Y_{3m}$ to its octupole moment. 
First we demonstrate that the octupole and
all higher moments vanish for Gd$^{3+}$ if one neglects the spin-orbit
interaction. 
The kinematic structure of the T,P-odd Hamiltonian  is, see, e.g. \cite{KL}
\begin{equation}
\label{hk}
   H_{TP} =a(r) {\bf s \cdot n}_r \; ,
\end{equation}
where $\bf s$ and ${\bf r}$ are the  spin and the position of the
electron, and ${\bf n}_r={\bf r}/r$ . 
An intermediate state $|n \rangle $ in (\ref{f6}) belongs to some
electronic configuration. We denote the configuration by ${\overline n}$. 
The states
within the configuration we numerate by an index $\alpha$, so
$|n \rangle =|{\overline n}_{\alpha} \rangle $. We also denote by $E_{\overline n}$
the average energy of the configuration.
In eq.(\ref{f6}) we first perform summation over $\alpha$ and
then over ${\overline n}$. 
With account of (\ref{hk}) this gives
\begin{equation}
\label{flm}
   f_{lm}(R) \propto \sum_{\overline n}
   \frac{\langle 0 |\delta(R-r) | {\overline n} \rangle 
   \langle {\overline n}| a(r)|0 \rangle}
   {E_0 - E_{\overline n}}
\langle 0| Y_{lm} s_i n_i |0\rangle
\end{equation}
Here we have used the closure relation
$\sum_{\alpha} |{\overline n}_{\alpha} \rangle \langle {\overline n}_{\alpha}|
= 1$, where 1 acts in the spin-angular space.
This relation is valid because of the separation of variables in the problem.
Without  the spin-orbit interaction,
the ground state of Gd$^{3+}$ reads
\begin{equation} \label{f0}
   |0 \rangle = |L=0\rangle |S= 7/2 \rangle,
\end{equation}
where $|L=0\rangle$ is the S-wave orbital state (half filled shell).
Hence we get from (\ref{flm})
\begin{equation}
   f_{lm} \propto \langle 0| Y_{lm} s_i n_i |0\rangle
=  \langle L=0 | Y_{lm} n_i | L=0 \rangle
   \langle S=7/2 | s_i|S=7/2 \rangle.
\end{equation}
It is clear that this equation gives zero for any $l$ higher than 1
because  one can combine $Y_{lm}$ with ${\bf n}$ into
total angular momentum zero only at $l=1$.
The spin-orbit interaction ${\bf l\cdot s}$
admixes states with nonzero $L$ to the ground state of Gd$^{3+}$.
The interaction contains the first power of orbital angular momentum ${\bf l}$,
so each order of the perturbation theory in the interaction changes the
selection rule by $\Delta l =\pm 1$. Hence, to get a nonzero octupole moment
of Gd$^{3+}$ one has to go to the second order in the spin-orbit interaction.
This is a very small effect and we do not consider it.

\newpage
Now we return to the energy correction (\ref{f4}).
Expanding $|\bf r - R|^{-1}$ in a series of spherical harmonics
\begin{eqnarray} \label{f5}
   \frac{1}{ | {\mathbf{r} - \mathbf{R}} | } =
   \sum_{l=0}^{\infty} \sum_{m=-l}^{l}
   \frac{4 \pi}{2l + 1}
   \frac{r^{l \phantom{+1}}_<}{r^{l+1}_>}
   Y_{lm}^*(\theta_r, \phi_r) Y_{lm}(\theta_{r'}, \phi_{r'})
   \\ \nonumber
   \text{where} \quad r_> = \max(r,R) \quad \text{and} \quad r_< =
   \min(r,R),
\end{eqnarray}
and keeping in mind that according to the previous discussion
$\delta\rho({\bf r})$ contains only the first harmonic, we transform
the energy correction (\ref{f4}) to the following form
\begin{eqnarray} \label{f7}
  \Delta \epsilon &=&
   e^2 \int d^3R \; q({\mathbf{R}}) \cos \theta_R
   \left[
      \frac{1}{R^2} \int_0^R d^3r\,
      \delta \rho({\mathbf{r}})r \cos \theta_r
      + R\int_R^{\infty} d^3r\;
      \frac{\delta \rho({\mathbf{r}})\cos \theta_r} {r^2}
      \right.
      \\ \nonumber
      & &
      \qquad  \qquad \qquad \qquad  \qquad \qquad \qquad \qquad \qquad
      \left.
      -\frac{1}{3R^2}
      \int_0^R d^3r\, \rho_0({\mathbf{r}})
      \int d^3r'\; \delta \rho({\mathbf{r'}})r' \cos \theta'_r
   \right]
\end{eqnarray}
This formula is similar to the standard one for the Schiff moment 
contribution to the T,P-odd energy correction, see Refs. \cite{SFK,KL}. 
For a smooth function $q(R)$ the eq.
(\ref{f7}) can be easily transformed to the standard form.
However the point is that in the present
case the density $q(R)$ is not a smooth function of $R$. Similar to the
standard case, eq. (\ref{f7}) gives zero $\Delta \epsilon$
if the external charge $q(R)$ is
localized outside of the Gd$^{3+}$ core. So we have to calculate the
probability for external electrons to penetrate inside the Gd$^{3+}$ core.
This calculation is performed in the following section.
Considering the integral over $\bf R$ in eq.~(\ref{f7}), we deduce
that only an asymmetric part in $q({\bf R})$ proportional to $\cos \theta_R$ 
can contribute to the energy correction. There is no such term for the 
lattice without deformation. Hence the nonzero $\Delta \epsilon$ is also 
related to the deformation of the lattice.

\section{Penetration of Oxygen electrons inside the Gd$^{3+}$ core, 
Deformation of the lattice.}

In the Garnet structure any Gd$^{3+}$ ion is surrounded by eight O$^{2-}$ 
ions organized in a dodecahedron structure (slightly distorted cube) 
\cite{GGG}. 
For the calculation we approximate this structure by a cube with
Gd$^{3+}$ in the middle. The Gd-O distance  used is $4.53a_B$,
where $a_B$ is the Bohr radius.
In the case of GdIG there are also uncompensated spins localized at
Fe$^{3+}$ ions. However the T,P-odd effect scales as
$Z^3$, and therefore the contribution of Fe is negligible compared to
that of Gd. An additional reason for suppression of the Fe contribution
is that the crystal radius of Fe$^{3+}$  is much smaller than that  
 of Gd$^{3+}$.
The quantity of interest is the charge density of  external electrons
$q(\bf r)$ inside the Gd$^{3+}$ ion. The electrons are coming from the
eight nearest O$^{2-}$ ions.

\begin{figure}[h]
   \vspace{20pt}
   \hspace{-35pt}
   \epsfxsize=5cm
   \centering\leavevmode\epsfbox{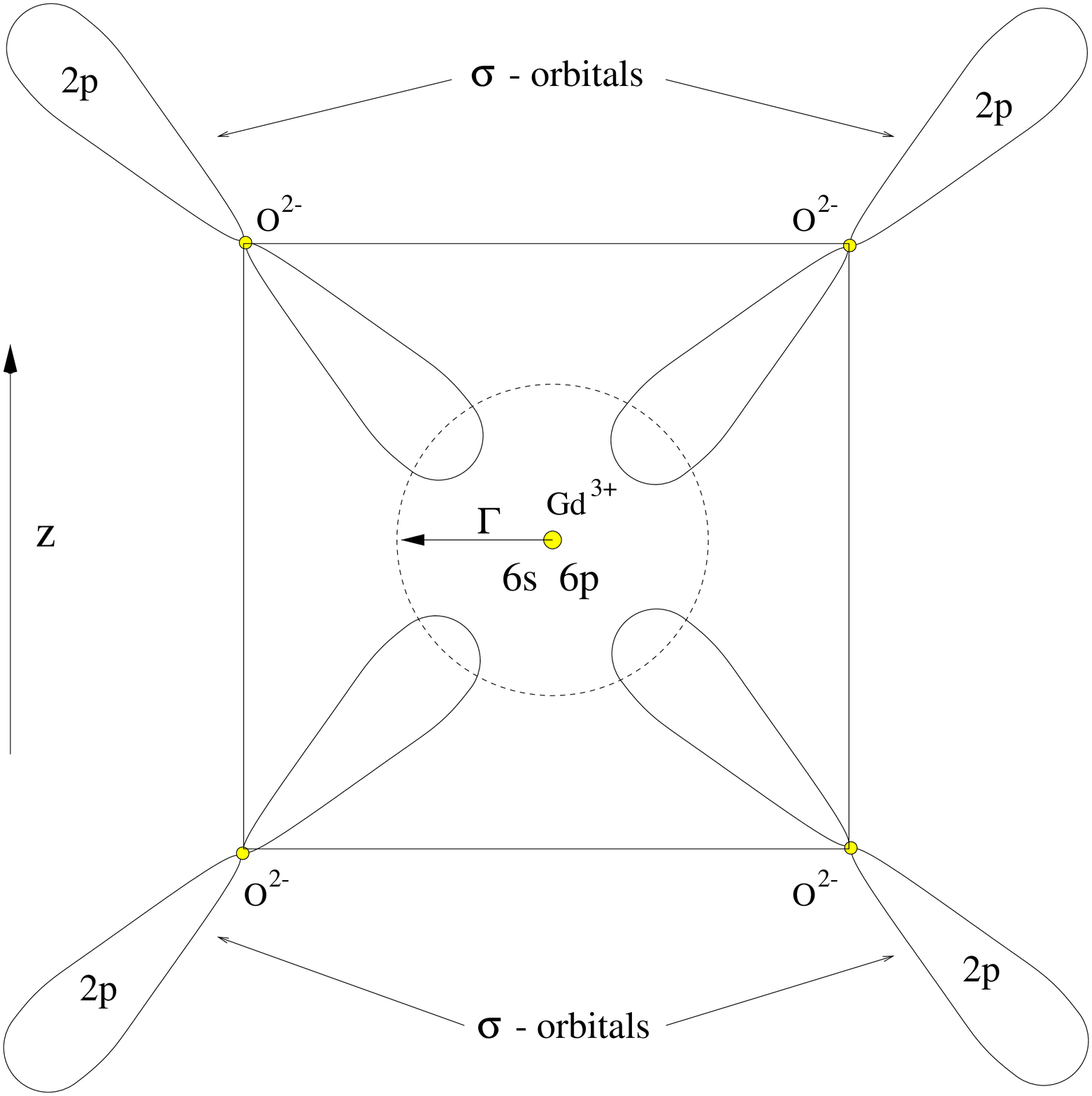}
   \vspace{8pt}
   \caption{\it {A schematic 2-dimensional picture for penetration of 
2p$_{\sigma}$-orbitals of O$^{2-}$ inside Gd$^{3+}$ }}
   \label{Fig1}
\end{figure}

The O$^{2-}$ ion has a closed shell configuration, with 3 double
occupied 2p-orbitals. One can show that  $2p_{\pi}$ orbitals do not
contribute to the effects we are interested in. So we only consider 
$2p_{\sigma}$-orbitals pointing towards Gd$^{3+}$ as schematically 
shown in the 2-dimensional picture in Fig.~\ref{Fig1}.  Hence there are  
16 electrons that can potentially penetrate inside Gd$^{3+}$.

\begin{figure}[h]
   \vspace{2pt}
   \hspace{-35pt}
   \epsfxsize=6cm
   \centering\leavevmode\epsfbox{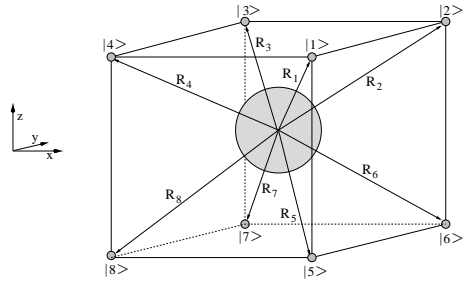}
   \vspace{8pt}
   \caption{\it {The matching sphere around Gd with the
   surrounding eight Oxygen neighbors.}}
   \label{Fig2}
\end{figure}

\newpage
First of all let us rewrite the eight $2p_{\sigma}$-orbitals in the basis
with definite symmetry with respect to the cubic structure.
The new basis reads
\begin{eqnarray} \label{f10}
\!\!\!\!\!\!
|S \rangle \quad\:\, &=&
(1/\sqrt8)%
\left( \;
|1 \rangle +\!|2 \rangle +\!|3 \rangle +\!|4 \rangle
+  \!|5 \rangle +\!|6 \rangle +\!|7 \rangle +\!|8 \rangle \;
\right)
\\ \nonumber
|P_x \rangle \;\;\:\,&=&
(1/\sqrt8)%
\left( \;
|1 \rangle +\!|2 \rangle -\!|3 \rangle -\!|4 \rangle
+  \!|5 \rangle +\!|6 \rangle -\!|7 \rangle -\!|8 \rangle \;
\right)
\\  \nonumber
|P_y \rangle \;\;\:\,&=&
(1/\sqrt8)%
\left( \;
|1 \rangle -\!|2 \rangle -\!|3 \rangle +\!|4
\rangle
+  \!|5 \rangle -\!|6 \rangle -\!|7 \rangle
+\!|8 \rangle \;
\right)
\\  \nonumber
|P_z \rangle \;\;\:\,&=&
(1/\sqrt8)%
\left( \;
  |1 \rangle +\!|2 \rangle +\!|3 \rangle +\!|4 \rangle 
  - \!|5 \rangle -\!|6 \rangle -\!|7 \rangle -\!|8 \rangle \;
\right)
\\ \nonumber
|D_{xy} \rangle \;&=&
(1/\sqrt8)%
\left( \;
|1 \rangle -\!|2 \rangle +\!|3 \rangle -\!|4 \rangle
+  \!|5 \rangle -\!|6 \rangle +\!|7 \rangle -\!|8 \rangle \;
\right)
\\ \nonumber
|D_{xz} \rangle \;&=&
(1/\sqrt8)%
\left( \;
|1 \rangle +\!|2 \rangle -\!|3 \rangle -\!|4 \rangle
-  \!|5 \rangle -\!|6 \rangle +\!|7 \rangle +\!|8 \rangle \;
\right)
\\ \nonumber
|D_{yz} \rangle \;&=&
(1/\sqrt8)%
\left( \;
|1 \rangle -\!|2 \rangle -\!|3 \rangle +\!|4 \rangle
-  \!|5 \rangle +\!|6 \rangle +\!|7 \rangle -\!|8 \rangle \;
\right)
\\ \nonumber
|F_{xyz} \rangle &=&
(1/\sqrt8)%
\left( \;
|1 \rangle -\!|2 \rangle +\!|3 \rangle -\!|4 \rangle
-  \!|5 \rangle +\!|6 \rangle -\!|7 \rangle +\!|8 \rangle \;
\right)
\end{eqnarray}
Here, for example $|1 \rangle $ denotes the $2p_{\sigma}$-orbital of the first
oxygen. The ions are numbered as in Fig.~\ref{Fig2}.
All of the states~(\ref{f10}) are doubly occupied.
As long as the electron is moving near the Oxygen ion the electron
wave function can be described as a $2p_{\sigma}$-orbital. However when 
the electron approaches the Gd$^{3+}$ ion its wave function is
strongly deformed by the potential of Gd$^{3+}$. 
To describe this effect we draw an imaginary sphere of radius $\Gamma$
around the Gd$^{3+}$ ion as  shown in Fig.~\ref{Fig1}.
We assume that outside the sphere the Oxygen potential dominates
and the wave is described as one of the (\ref{f10}) orbitals.
However inside the sphere the Gd$^{3+}$ potential dominates and hence
the same orbital is described as s-, p-, d-, or f-orbital of Gd.
We are interested only in s- and p- orbitals that correspond to
$|S \rangle $, $|P_x \rangle $, $|P_y \rangle $, and $|P_z \rangle$ 
states from (\ref{f10}).
At $r < \Gamma$ we will use the 6s and 6p orbitals of Gd$^{1+}$ as the basis.
 We can also choose 7s and 7p or 8s and 8p,  the
final result is independent of the choice because at $r < \Gamma$ all
these wave functions are proportional to each other.
\noindent
Thus at $r < \Gamma$,  we get
\begin{equation} \label{f11}
   |S\rangle \to \beta_S |6s\rangle, 
   \qquad \qquad 
   |P_z\rangle \to \beta_P |6p_z\rangle,
   \qquad \qquad 
   |P_x\rangle \to \beta_P |6p_z\rangle,
   \qquad \qquad 
   |P_y\rangle \to \beta_P |6p_z\rangle.
\end{equation}
To find the coefficients $\beta_S$ and $\beta_P$ we match (\ref{f11}) with
(\ref{f10}) at the surface of the sphere with radius $r=\Gamma$ using the
following projection
\begin{equation} \label{f12}
   \langle \,
   u(\Gamma,\phi,\theta) \,|\,
   v(\Gamma,\phi,\theta) \,
   \rangle_{_{\Gamma}}
   =
   \int_0^{2 \pi} \!\!\!  d\phi  \!
   \int_0^{\pi} \!\! d\theta \, \sin \theta \,
   u^{*}(\Gamma,\phi,\theta) \,
   v    (\Gamma,\phi,\theta).
\end{equation}
Hence
\begin{equation} \label{f13}
   \beta_S =
   \frac{\langle 6s \,|\, S  \, \rangle_{_{\Gamma}} }
   {\langle 6s \,|\, 6s \, \rangle_{_{\Gamma}}}
   \qquad \qquad \qquad
   \beta_P =
   \frac{\langle 6p_z \,|\, P_z  \, \rangle_{_{\Gamma}} }
   {\langle 6p_z \,|\, 6p_z \, \rangle_{_{\Gamma}}}.
\end{equation}
This is a simple and reliable method to calculate the coefficients
$\beta_S$ and $\beta_P$. It has been previously used for calculation
of the nuclear quadrupole resonance frequencies in La$_2$CuO$_4$, see
Ref. \cite{FS}. In that case the accuracy of the method was verified by
experimental data and proved to be  $\sim$10\%.
For numerical calculations with (\ref{f12}),(\ref{f13}) we have used
2p-wave-functions of O$^{2-}$ calculated previously in \cite{FS}. The 6s and 6p
wave functions of Gd$^{1+}$ have been obtained by the relativistic 
Hartree-Fock method for configuration $1s^2....4f^76s6p$. In this calculation
we average over polarizations of the unclosed shells. It means that $4f^7$
is taken as $4f^{14}$ with 50\% population of each orbital, similarly
6s is taken as $6s^2$ with 50\% population and 6p is taken as $6p^6$ with
1/6=16.6\% population. There is also the fine structure effect: the difference
between $6p_{1/2}$ and $6p_{3/2}$ states. However this effect is small and has
been neglected in the calculation of $\beta_P$.
The values of $\beta_S$ and $\beta_P$ calculated with (\ref{f12}),(\ref{f13})
for different choices of the matching radius $\Gamma$ are presented in 
Table 1. The coefficients are not sensitive to the value of $\Gamma$ and
this confirms validity of the method.
\begin{table}[t]
\label{t1}
\begin{tabular}{r@{.}l r@{.}l r@{.}l r@{.}l r@{.}l r@{.}l r@{.}l r@{.}l  }
   \multicolumn{2}{c}{$\Gamma\; [a_B]$} &
   \multicolumn{2}{c}{$\beta_{S}$} &
   \multicolumn{2}{c}{$\beta_{P}$} &
   \multicolumn{2}{c}{$\beta'_S$} &
   \multicolumn{2}{c}{$\beta'_P$} &
   \multicolumn{2}{c}{$P_{s}$} &
   \multicolumn{2}{c}{$P_{p}$} &
   \multicolumn{2}{c}{$\langle N_{2p} \rangle$} \\
   \hline
   2&2 & -0&448 & 0&829 & 0&386 & -0&210 & 1&7\% & 2&7\% & 0&19  \\
   2&3 & -0&446 & 0&754 & 0&345 & -0&204 & 2&1\% & 2&7\% & 0&20  \\
   2&4 & -0&447 & 0&704 & 0&316 & -0&200 & 2&4\% & 2&7\% & 0&21  \\
   \hline
   2&5 & -0&448 & 0&650 & 0&285 & -0&196 & 3&2\% & 2&8\% & 0&23  \\
   \hline
   2&6 & -0&461 & 0&639 & 0&273 & -0&196 & 4&1\% & 3&3\% & 0&28  \\
   2&7 & -0&478 & 0&634 & 0&263 & -0&197 & 5&0\% & 3&6\% & 0&31  \\
   2&8 & -0&499 & 0&637 & 0&254 & -0&198 & 6&2\% & 4&5\% & 0&39  \\
\end{tabular}
\vspace{10pt}
\caption{The coefficients $\beta_S$, $\beta_S'$, 
$\beta_P$, and $\beta_P'$ for expansion of Oxygen 2p-orbitals in
terms of Gd$^{1+}$ 6s- and 6p-wave-functions, see eqs. (\ref{f18}),(\ref{f11}).
The coefficients are presented for different values of the radius of 
the matching sphere $\Gamma$. $P_s$ and $P_p$ represent the probability
for the corresponding Oxygen electron to penetrate inside the matching
sphere. $\langle N_{2p}\rangle$ is the average number of Oxygen 2p-electrons
inside the matching sphere.}
\end{table}
\noindent
The  calculation assumes that the 2p-wave-functions of O$^{2-}$ are
``rigid'', i.e. at $r> \Gamma$ they are not influenced by Gd. This
can be true only if the total probability for an Oxygen electron to penetrate
inside the matching sphere, $r < \Gamma$, is small. To check this we calculate
the probabilities
\begin{eqnarray} \label{f14}
   P_{s} &=&
   \beta_S^2 \int_0^{\Gamma} d^3r\, 
   |\psi_{6s}({\bf r})|^2 ,
   \\
   P_{p} &=&
   \beta_P^2 \int_0^{\Gamma} d^3r\,
   |\psi_{6p}({\bf r})|^2.
\end{eqnarray}
Numerical values of $P_{s}$ and $P_{p}$ are presented in Table 1, 
they are really very small.
The average number of Oxygen electrons penetrating inside the matching sphere,
\begin{equation} \label{f15}
   \langle N_{2p} \rangle = 2(P_{s} + 3P_{p}),
\end{equation}
listed in Table 1 is also rather small.
Thus, the external charge density with respect to Gd$^{3+}$  inside the
matching sphere is given by
\begin{equation} \label{f16}
  q({\bf r}) =q_0({\bf r})=
  2\beta_S^2 |\psi_{6s}({\bf r})|^2 +
  2\beta_P^2 \left( 
|\psi_{6p_x}({\bf r})|^2+
|\psi_{6p_y}({\bf r})|^2+
|\psi_{6p_z}({\bf r})|^2
  \right)
\end{equation}
The coefficient 2 is due to the double occupancy (spin up, down) of 
each orbital.

\begin{figure}[h]
   \vspace{2pt}
   \hspace{-35pt}
   \epsfxsize=5cm
   \centering\leavevmode\epsfbox{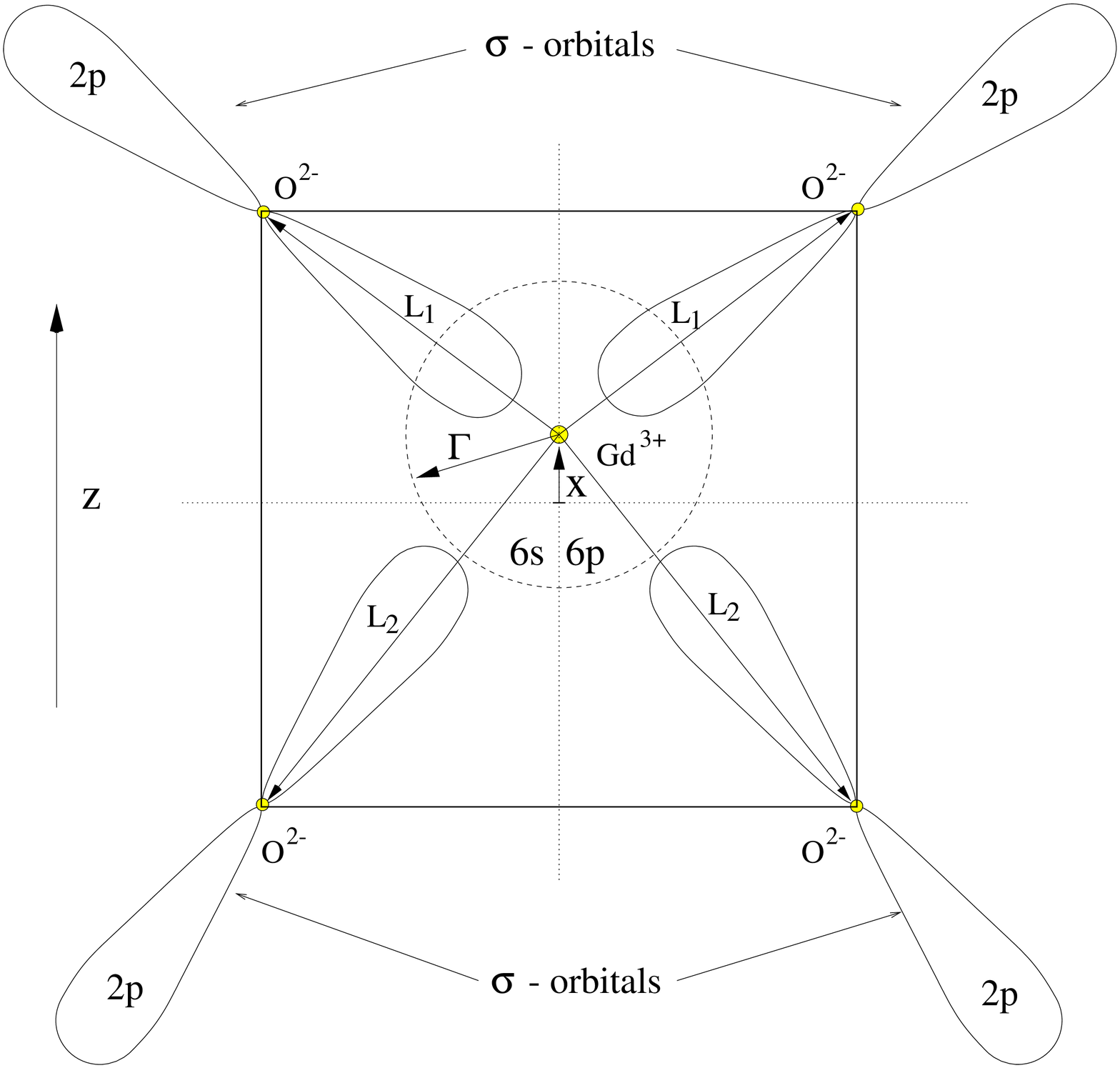}
   \vspace{8pt}
   \caption{\it {A schematic 2-dimensional picture for penetration of 
2p$_{\sigma}$-orbitals of O$^{2-}$ inside the shifted Gd$^{3+}$}}
   \label{Fig3}
\end{figure}

It has been pointed out in the previous section that the charge density 
(\ref{f16}) is insufficient
to generate a T,P-odd effect because $q_0({\bf r})$ is symmetric with
respect to reflection ${\bf r} \to -{\bf r}$. To have an asymmetric
part $\delta q({\bf r})$ one needs to consider a deformation of the lattice.
We will assume that Gd is shifted along the z-axis with respect to the
environment by a small displacement X, for definition of axes see 
Fig.~\ref{Fig2}. The displacement is shown schematically  
in the 2-dimensional picture in Fig.~\ref{Fig3}.
In the general case the displacement can have an arbitrary direction.
However we do not expect that the final answer will be very sensitive
to the direction. Therefore we consider the simplest geometry:
the displacement along the axis of the cube.
\noindent
It is clear that outside the matching sphere nothing is changed: eqs. 
(\ref{f10}) describe the wave functions. 
\begin{table}
\end{table}
\noindent
However inside the sphere the wave functions 
are changed compared to (\ref{f11})
\begin{equation} \label{f17}
   |S\rangle 
   \to \beta_{S} |6s\rangle + \beta^{'}_{S}
   \frac{X}{a_B} |6p_z\rangle
   \qquad \qquad
   |P_z\rangle \to \beta_{P}
   |6p_z\rangle + \beta^{'}_{P}
  \frac{X}{a_B} |6s\rangle.
\end{equation}
We present only the orbitals that give a nonzero contribution to the
T,P-odd effect. To have dimensionless coefficients $\beta'_S$ and
$\beta'_P$ we divide the displacement by the Bohr radius.
To calculate the coefficients $\beta'_S$ and $\beta'_P$ we use the same
matching procedure (\ref{f12})
\begin{equation} \label{f18}
   \beta'_S 
   =   \left.
   \frac{{\partial }}{\partial X}
   \frac{\langle 6p_z | S \rangle_{_{\Gamma}}}
          {\langle 6p_z | 6p_z \rangle_{_{\Gamma}}}
   \right|_{X=0}
\qquad  \qquad
   \beta'_P
   =   \left.
   \frac{{\partial }}{\partial X}
   \frac{\langle 6s | P_z \rangle_{_{\Gamma}}}
          {\langle 6s | 6s \rangle_{_{\Gamma}}}
   \right|_{X=0}
\end{equation}
The values of $\beta'_S$ and $\beta'_P$ for different values of the
radius of the matching sphere $\Gamma$ are presented in Table 1.

As one can see from Table 1 the values for $\beta_S$,
$\beta_P$, $\beta'_S$ and $\beta'_P$ are almost independent of the
choice $\Gamma$ in the interval $2.2a_B - 2.8a_B$, this is the
dual description region. Existence of this region confirms validity of
our approach. For all further calculations we will use
$\Gamma = 2.5a_B$. 
Using eqs.~(\ref{f17}) we immediately obtain the
asymmetric term for the charge density,
%
\begin{equation} \label{f19}
\delta q ({\bf r})
=
4(\beta_S \beta'_S + \beta_P \beta'_P)
\frac{X}{a_B}  
\psi_{6s}({\bf r})\psi_{6p_z}({\bf r})
\end{equation}
In the above calculation we have neglected relativistic effects.
This is justified as far as the calculation of the coefficients $\beta$ is
concerned. However for the behavior of $\delta q({\bf r})$ near the Gd
nucleus the relativistic effects are very important.
Therefore we have to modify (\ref{f19}) to account for this effect.
Fortunately this modification is obvious and immediately follows from
decomposition of $6p_z$ in terms of $6p_{1/2}$ and $6p_{3/2}$.
This gives
\begin{equation} \label{f20}
   \delta q({\bf r}) \to 
   4(\beta_S \beta'_S + \beta_P \beta'_P)
   \frac{X}{a_B}
   \left(
      \sqrt{\frac{2}{3}} u_s^\dagger u_{p_{3/2}}
      -\sqrt{\frac{1}{3}} u_s^\dagger u_{p_{1/2}}
   \right)
\end{equation}
where
\begin{equation} \label{f21}
   u_{p_{1\!/\!2}} 
   =
   \left(
   \begin{array}{c}
      R_{p_{1\! / \!2}} \Omega_{p_{1\!/\!2}} \\
      i\tilde{R}_{p_{1\! / \!2}}
      \tilde{\Omega}_{p_{1\!/\!2}}
   \end{array}
   \right),
   \qquad
   u_{p_{3\!/\!2}} 
   =
   \left(
   \begin{array}{c}
      R_{p_{3\! / \!2}}
      \Omega_{p_{3\!/\!2}} \\
      i\tilde{R}_{p_{3\! / \!2}}
      \tilde{\Omega}_{p_{3\!/\!2}}
   \end{array}
   \right),
   \qquad
   u_s
   =
   \left(
   \begin{array}{c}
      R_s \Omega_s \\
      i\tilde{R}_s \tilde{\Omega}_s
   \end{array}
   \right),
\end{equation}
are the Dirac wave functions of  $6p_{1/2}$, $6p_{3/2}$, and $6s_{1/2}$
states. $\Omega$ and $\tilde{\Omega}=-({\bf \sigma \cdot n})\Omega$
are the usual two-component spherical spinors \cite{BLP}; $R$ and $\tilde R$ 
are upper and lower radial wave functions that have been calculated
using the Relativistic Hartree-Fock method.

\section{Relation between the lattice deformation and the T,P-odd energy 
correction induced by the electron EDM}

\begin{figure}[h]
\vspace{2pt}
\hspace{-35pt}
\epsfxsize=8cm
\centering\leavevmode\epsfbox{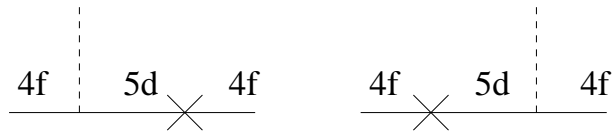}
\vspace{8pt}
\caption{\it {Leading contribution to Gd$^{3+}$ EDM.
The dashed line denotes the dipole moment $ez=er\cos\theta$, and the
cross denotes the T,P-odd  interaction $V_d^r$, eq (\ref{Vd1}).}}
\label{Figd}
\end{figure}

The T,P-odd interaction of the electron EDM with the electric field 
of the Gd nucleus ${\bf E}$ is $V_d=-d_e\gamma_0{\bf \Sigma\cdot E}$, and 
account of the Schiff theorem for electronic degrees of freedom reduces it 
to the form, see e.g. Ref. \cite{KL}
\begin{equation}
\label{Vd1}
V_d\to V_d^r=-d_e(\gamma_0-1){\bf \Sigma\cdot E},
\end{equation}
where  $\gamma_0$ and ${\bf \Sigma}=\gamma_0\gamma_5{\bf \gamma}$ are
Dirac $\gamma$-matrices.
According to \cite{Stefan} the EDM of Gd$^{3+}$ ion is saturated by 
4f-5d mixing
\begin{equation} \label{f22}
   d_a =K d_e =
   e \int_0^{\infty} d^3r\, \delta \rho({\bf r}) r\cos \theta_r
   \approx 
   2e \sum_{m=-2}^2
   \frac{\langle 4f_m |r \cos \theta_r | 5d_m\rangle
   \langle 5d_m | V_d^r| 4f_m\rangle}
   {E_{4f} - E_{5d} }.
\end{equation}
Diagrammatically it is shown in Fig.~\ref{Figd}
\noindent
The same 4f-5d virtual transitions saturate the T,P-odd correction to 
energy (\ref{f7}), and taking into account  eq. (\ref{f20}) the correction
(\ref{f7}) can be rewritten as
\begin{eqnarray} \label{f24}
   \Delta \epsilon 
   &=& 
   \frac{4e^2}{\sqrt{3}}
   (\beta_S \beta'_S + \beta_P \beta'_P)
    \frac{X}{a_B}
   2\!\! \sum_{m=-3}^{3}\!\!
   \frac{\langle Y_{3m}|\cos\theta|Y_{4m}\rangle
   \langle 5d_m|V_d^r| 4f_m \rangle} {E_{4f} - E_{5d}}
   \left(
      I_1 + I_2 + I_3
   \right)\nonumber\\
   &=&\frac{4e}{\sqrt{3}}
   (\beta_S \beta'_S + \beta_P \beta'_P)
    \frac{X}{a_B}{{K d_e}\over{\langle r_{fd}\rangle }}
\left( I_1 + I_2 + I_3   \right).
\end{eqnarray}
where,
\begin{eqnarray}  \label{f25}
  && I_1 =
   \int_0^{\Gamma} \!\! dR\,
   \left(
      \frac{1}{3}(R_sR_{p_{1/2}} + \tilde R_s \tilde R_{p_{1/2}} ) +
      \frac{2}{3}(R_sR_{p_{3/2}} + \tilde R_s \tilde R_{p_{3/2}} )
   \right)
   \! \int_0^R \!\!\! dr\;r^3 R_{4f}(r) R_{5d}(r),\\ 
  && I_2 =
   \int_0^{\Gamma} \!\! dR\,R^3\,
   \left(
      \frac{1}{3}(R_sR_{p_{1/2}} + \tilde R_s \tilde R_{p_{1/2}} ) +
      \frac{2}{3}(R_sR_{p_{3/2}} + \tilde R_s \tilde R_{p_{3/2}} )
   \right)
   \int_R^{\Gamma} \!\!dr\; R_{4f}(r) R_{5d}(r),\nonumber\\
  && I_3 =
   -\frac{1}{3}
   \int_0^{\Gamma} \!\! dR\,
   \left(
      \frac{1}{3}(R_sR_{p_{1/2}} + \tilde R_s \tilde R_{p_{1/2}} ) +
      \frac{2}{3}(R_sR_{p_{3/2}} + \tilde R_s \tilde R_{p_{3/2}} )
   \right)
   \int_0^R \!\! d^3r \, \rho_0({\bf r})
   \! \int_0^{\Gamma} \!\!\!dr'\;{r'}^3 R_{4f}(r') R_{5d}(r'),\nonumber\\
  && \langle r_{fd}\rangle =
 \int_0^{\Gamma} \!\!\!dr'\;{r'}^3 R_{4f}(r') R_{5d}(r'),\nonumber
\end{eqnarray}
In the integrals
we have replaced the upper limit of integration by $\Gamma=2.5a_B$,
because eq. (\ref{f20}) is valid only at $r < \Gamma$.
The final result is not sensitive to the upper limit because these integrals
 are convergent at smaller distances.
The numerical values  obtained with Hartree-Fock wave functions are
\begin{equation} \label{f28}
   I_1 = -4.1\times 10^{-3}\;a_B^{-1}, \qquad
   I_2 = -5.3\times 10^{-3}\;a_B^{-1}, \qquad
   I_3 = -7.5\times 10^{-3}\;a_B^{-1}, \qquad
  \langle r_{fd}\rangle =0.55a_B.
\end{equation}
Substituting these values, as well as the value of the coefficient
$K$ from (\ref{K}) and the coefficients $\beta$ from Table 1.
in eq. (\ref{f24}) we find the T,P-odd energy correction per  Gd ion
as a function of the ion displacement with respect to surrounding Oxygen ions.
\begin{equation} \label{f29}
   \Delta \epsilon(X)
   =-A \frac{e}{a_B^2} \frac{X}{a_B} d_e
\to -A \frac{e\ d_e}{S\ a_B^3} ({\bf X\cdot S})
   \qquad
   \text{with}
   \quad A = 0.11\;.
\end{equation}
The vector form is valid for an arbitrary direction of the displacement 
${\bf X}$ and an arbitrary direction of the Gd$^{3+}$ polarization ${\bf S}/S$.
The integrals (\ref{f28}) give $A=0.09$, however they only account for the
contribution of 2p-orbitals of Oxygen. We know \cite{FS} that there is
also a contribution of 2s-orbitals of Oxygen that is 20\%-25\% of the
2p-contribution. This is why we take $A=0.11$.

\begin{figure}[h]
\vspace{2pt}
\hspace{-35pt}
\epsfxsize=6cm
\centering\leavevmode\epsfbox{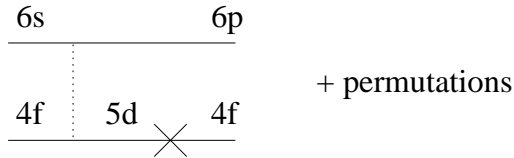}
\vspace{8pt}
\caption{\it {Direct contribution to the T,P-odd energy shift.
The dotted line denotes the Coulomb electron-electron interaction, and the
cross denotes the T,P-odd  interaction $V_d^r$, eq (\ref{Vd1}).}}
\label{fig12}
\end{figure}

\begin{figure}[h]
\vspace{2pt}
\hspace{-35pt}
\epsfxsize=10cm
\centering\leavevmode\epsfbox{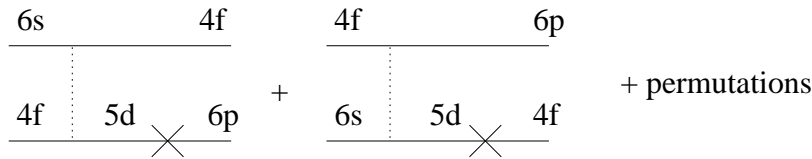}
\vspace{8pt}
\caption{\it {Exchange contribution to the T,P-odd energy shift.
The dotted line denotes the Coulomb electron-electron interaction, and the
cross denotes the T,P-odd  interaction $V_d^r$, eq (\ref{Vd1}).}}
\label{figex}
\end{figure}

The integral $I_3$ corresponds to the compensating term in (\ref{f7})
and the contributions corresponding to the integrals $I_1$ and $I_2$
can be represented diagrammatically as  shown in Fig.\ref{fig12}
\noindent
This is what we have taken into account and these are direct diagrams only. 
In principle there are also exchange diagrams shown schematically in 
Fig. \ref{figex} that we have neglected.
\noindent
The exchange diagrams have higher angular momenta of the Coulomb quantum than
the direct ones, the diagrams in Fig.\ref{figex} have $k=2,3,4$, while the
direct diagram in Fig.\ref{fig12} has $k=1$.
Therefore it is unlikely that the exchange contributions 
can be comparable with direct ones. Nevertheless, at some later stage, 
these contributions have to be calculated as well.

\section{T,P-odd voltage across a magnetically polarized sample
of GdIG}
In this section we consider a GdIG sample magnetically polarized along some
axis. Then according to eq. (\ref{f29}) each Gd ion can gain energy
from a small distortion of the lattice. However the lattice has stiffness
and therefore the total variation of energy per Gd ion as a function of the
displacement $X$ is of the form
\begin{equation}\label{f30}
   \Delta \epsilon_T =
   \frac{1}
   {2}
   K_{el} \, X^2
   -   A   \frac{e}{a_B^2}
   \frac{X}{a_B} d_e,
\end{equation}
where $K_{el}$ is the effective elastic constant per Gd ion.
Minimizing (\ref{f30}) with respect to the lattice deformation $X$
we find the deformation induced by the electron EDM.
\begin{equation}\label{f32}
   X =
    \frac{A\,e}{ K_{el} a_B^3} d_e.
\end{equation}
To calculate the deformation we need to know the value of $K_{el}$. 
The elastic constant is related to optic phonons,
and can be found from the spectra of the phonons. However we will
use a simpler method based on the known static dielectric constant of GdIG,
$\epsilon\approx 15$ \cite{epGIG}.
Consider some external electric field applied to the GdIG sample.
According to standard relations
\begin{equation}
\label{ed}
{\bf D}=\epsilon {\bf E}={\bf E}+4\pi{\bf P},
\end{equation}
where $\bf E$ is the average electric field in the sample and $\bf P$ the 
polarization per unit volume. Assuming simple cubic structure
the local field acting on each ion is \cite{Kit}
\begin{equation}
\label{loc}
{\bf E_l} = {\bf E} + {{4 \pi}\over{3}} 
{\bf P}=4\pi{\bf P}\left({1\over{\epsilon-1}}+
{1\over{3}}\right)
\end{equation}
This gives the displacement of the Gd$^{3+}$ ion in the field
\begin{equation}
\label{x1}
x_1={{3eE_l}\over{K_{el}}}={{12\pi e}\over{K_{el}}}P\left({1\over{\epsilon-1}}+
{1\over{3}}\right)
\end{equation}
Similarly one can find the displacement of the Fe$^{3+}$ ion $x_2$ in terms 
of the corresponding elastic constant $K_{el}'$. On the other hand,
the polarization per unit volume by definition is 
$P=3e x_1 n_{Gd}+5e x_2 n_{Gd}$. Altogether this gives the following result
for $K_{el}$.
\begin{equation} \label{f31}
   K_{el}  =  96 \pi e^2 n_{Gd}
   \left(
   \frac{1}{\epsilon -1}
   +\frac{1}{3}
   \right)\left({3\over{8}}+{{5K_{el}}\over{8K'_{el}}}\right).
\end{equation}
We recall that $n_{Gd}=1.235\times 10^{22}\; cm^{-3}$ is the number density 
of Gd in GdIG. For further estimates we will assume that $K_{el}=K_{el}'$.
There is no reason for these constants to differ substantially, and
moreover eq. (\ref{f31}) is not that sensitive to the ratio $K_{el}/K_{el}'$.

Coming back to the situation without any external electric field and using 
eqs. (\ref{f32}) and (\ref{f31}) we find the electric polarization 
$P=3eXn_{Gd}$ and hence the electric field inside the sample
\begin{equation}
\label{EE}
E = -4\pi P=-{{A}\over{8\left({1\over{\epsilon-1}}+{1\over{3}}\right)}}
\cdot {{d_e}\over{a_B^3}}=-0.54\times 10^{-10}V/cm.
\end{equation}
This numerical value corresponds to the current upper limit on the electron
EDM (\ref{Tl}).
For a 10 cm sample it gives a voltage $ \Delta V = 0.54\times 10^{-9}V$.
This is smaller than the naive estimate (\ref{naive}) and this is
suppression due to the Schiff theorem. Fortunately this suppression is
not  strong.

\section{T,P-odd magnetization of GGG in the external electric field}
Following the suggestion of Lamoreaux \cite{Lam} we consider now a different 
kind of 
experiment: an external electric field is applied to the sample and this
leads to the macroscopic magnetization.
Equation (\ref{x1}) gives the Gd shift due to the electric field.
\begin{equation}
   X=\frac{(\epsilon -1)}{32 \pi e n_{Gd}} E
\end{equation}
Substituting this into eq. (\ref{f29}) we find the energy shift of the
Gd$^{3+}$ ion
\begin{equation}
\label{ddee}
   \Delta \epsilon = -\frac{A(\epsilon -1)}
   {32 \pi (a_B^3 n_{Gd})}d_e {{({\bf E\cdot S})}\over{S}}\to
2.8\times 10^{-22}eV.
\end{equation}
The numerical estimate corresponds to $E=10kV/cm$, the current limitation 
on  $d_e$, (\ref{Tl}), and the dielectric constant $\epsilon \approx 30$ for 
GGG \cite{epGIG}.
It is interesting that in spite of the Schiff theorem the result 
(\ref{ddee}) is larger than the naive estimate (\ref{n2}).

The energy shift (\ref{ddee}) leads to the macroscopic magnetization of the
sample. The magnetization depends also on temperature and internal magnetic
interactions in the compound. We do not discuss these points here.
According to estimates \cite{Lam} the magnetization due to the energy
shift $\sim 10^{-22}eV$ can be measured and moreover the prospects for 
improvement of sensitivity are very good.

\section{T,P-odd voltage and magnetization due to the Nuclear Schiff
Moment}
The effects considered above are due to the electron EDM, so they are
sensitive to T,P-violation in the lepton sector.
Similar effects are generated by the Nuclear Schiff Moment (NSM), so they are
also sensitive to T,P-violation in the hadron sector.
We begin from consideration of the same compounds GdIG and GGG, assuming
only one Gd isotope. This must be an odd isotope, say $^{155}$Gd.
The NSM ${\bf S}_N$ is a T,P-odd vector moment that is a property of the
nucleus, see Refs. \cite{SFK,KL}
\begin{equation}
\label{S}
{\bf S}_N= S_N{{\bf I}\over{I}},
\end{equation}
where {\bf I} is nuclear spin. In previous sections we denoted
by {\bf S} the total electron spin, note that it has nothing in common
with the NSM ${\bf S}_N$. 
Due to the Schiff moment there is the T,P-odd interaction between electron
and nucleus. This is the contact interaction, therefore the matrix
elements of the interaction are nonzero only for s- and p-electrons.
These matrix elements have been calculated in Ref. \cite{SFK}, see also
\cite{KL}
\begin{equation}
\label{sp}
\langle s|H_S|p\rangle={{Z^2eR}\over{\pi a_B^4(\nu_s\nu_p)^{3/2}}} {\bf S}_N
\langle\Omega_s|{\bf n}|\Omega_p\rangle.
\end{equation}
Here $\Omega_s$ and $\Omega_p$ are spherical spinors corresponding to the
s- and p-states; ${\bf n}$ is the unit vector; $\nu_s$ and $\nu_p$ are
principal effective quantum numbers; and $R$ is the relativistic factor
that is different for $p_{1/2}$ and $p_{3/2}$ electrons
\begin{eqnarray}
\label{rf}
&&R_{1/2}\approx {{4\gamma_{1/2}x_0^{2\gamma_{1/2}-2}}\over
{[\Gamma(2\gamma_{1/2}+1)]^2}},\\
&&R_{3/2}\approx {{48\gamma_{1/2}x_0^{\gamma_{1/2}+\gamma_{3/2}-3}}\over
{\Gamma(2\gamma_{1/2}+1)\Gamma(2\gamma_{3/2}+1)}}.\nonumber
\end{eqnarray}
Here $\gamma_{1/2}=\sqrt{1-Z^2\alpha^2}$, $\gamma_{3/2}=\sqrt{4-Z^2\alpha^2}$,
$\alpha$ is the fine structure constant; $\Gamma(x)$ is the usual 
$\gamma$-function; and $x_0=(2Zr_0/a_B)$, where $r_0$ is the nuclear radius.
For Gd, $R_{1/2}=3.56$ and $R_{3/2}=2.99$.
Assuming that the nucleus is completely polarized along the z-axis, and taking 
also $|p \rangle \to |p_z \rangle $ we find from (\ref{sp})
\begin{equation}
\label{spz}
\langle s|H_S|p_z\rangle={{S_NZ^2e}\over{\pi \sqrt{3}a_B^4(\nu_s\nu_p)^{3/2}}}
\left({1\over{3}}R_{1/2}+{2\over{3}}R_{3/2}\right)
\end{equation}
According to (\ref{f17}) lattice deformation $X$ leads to the mixing
of 6s- and 6p-orbitals. Hence the matrix element of $H_S$ over the
electronic states of the deformed lattice is equal to
\begin{equation}
\label{spz1}
\langle H_S\rangle=
{{4}\over{\pi\sqrt{3}}}(\beta_S\beta_S'+\beta_P\beta_P'){{X}\over{a_B}}
{{S_NZ^2e}\over{a_B^4(\nu_s\nu_p)^{3/2}}}
\left({1\over{3}}R_{1/2}+{2\over{3}}R_{3/2}\right).
\end{equation}
The effective principal numbers for 6s- and 6p-orbitals are:
$\nu_{6s}=1.24$, $\nu_{6p_{1/2}}=1.46$, and $\nu_{6p_{3/2}}=1.49$,
see comment \cite{c}.
Substituting these values, together with coefficients $\beta$ from Table 1
and all the other parameters we find the T,P-odd energy correction
due to the NSM
\begin{equation} \label{f40}
   \Delta \epsilon(X)
   =-B \frac{e}{a_B^4} \frac{X}{a_B} S_N
\to -B \frac{e\ S_N}{I \ a_B^5} ({\bf X\cdot I})
   \qquad
   \text{with}
   \quad B = 1.2\times 10^3\;.
\end{equation}
The vector form is valid for an arbitrary direction of the displacement 
${\bf X}$ and an arbitrary direction of the Gd nuclear spin ${\bf I}$.
Equation (\ref{f40}) for the energy shift due to NSM is similar to eq.
(\ref{f29}) for the energy shift due to the electron EDM.
We stress once more that $S_N$ in (\ref{f40}) denotes the NSM while $S$
in (\ref{f29}) denotes the electron spin.
In (\ref{f40}) we have also  taken into account 20\% contribution
from the Oxygen 2s-electrons, see comment after eq. (\ref{f29}).

Next we follow the same path as we did for the effects due to the
electron EDM. First consider a sample with fully polarized nuclear
spins, no external electric field. Then, repeating the calculations 
performed in Section VI we
find the following electric field induced inside the solid
\begin{equation}
\label{ES}
E =-{{B}\over{8\left({1\over{\epsilon-1}}+{1\over{3}}\right)}}
\cdot {{S_N}\over{a_B^5}}=-0.9\times 10^{-12}V/cm.
\end{equation}
The numerical value corresponds to the current upper limit on the NSM
of $^{199}$Hg (\ref{Hg}).

If the external electric field is applied to the solid then repeating the
calculations from Section VII we find the energy shift per Gd depending
on the orientation of Gd nucleus
\begin{equation}
\label{ddss}
   \Delta \epsilon = -\frac{B(\epsilon -1)}
   {32 \pi (a_B^3 n_{Gd})}\cdot {S\over{a_B^2}}\cdot 
{{({\bf E\cdot I})}\over{I}}\to
-0.47\times 10^{-23}eV.
\end{equation}
The numerical estimate corresponds to $E=10kV/cm$, the current limitation 
on  NSM of $^{199}$Hg (\ref{Hg}), and the dielectric constant 
$\epsilon \approx 30$ for GGG \cite{epGIG}.

\section{Accuracy of the calculations}
The effects related to the Nuclear Schiff moment are sensitive only to 
s- and p-orbitals of Gd. There are three sources of possible corrections
to the results presented in eqs. (\ref{f40}), (\ref{ES}), and (\ref{ddss}): 
1)interatomic many-body correlations, 2)inaccuracy 
of the used matching procedure for calculation of the expansion coefficients 
$\beta$, 3)deviation of the effective elastic constant $K_{el}$ for Gd
from that for Fe. Usually for s- and p-orbitals the correction due to the 
interatomic correlations does not exceed 20\%, see e.g. \cite{KL}.
As we already mentioned above the matching procedure has been previously used 
for calculation of the nuclear quadrupole resonance frequencies in 
La$_2$CuO$_4$, see Ref. \cite{FS}. In that case the accuracy of the procedure
was verified by experimental data and proved to be  about 10\%.
We believe that this estimate is also valid for the present case.
Finally, even in the unlikely case of 40\% difference in elastic
constants $K_{el}$ and $K_{el}'$ the variation in (\ref{ES}) and (\ref{ddss})
is 20\%. All in all this gives a 20\%-30\% estimate for a possible inaccuracy
in (\ref{f40}), (\ref{ES}), and (\ref{ddss}).

Situation with effects induced by the electron EDM is more complex.
The mechanisms 1-3  for possible corrections contribute here as well. The 
interatomic many-body correlations have been considered in 
Ref. \cite{Stefan}. They contribute at the
level 20\%, but the most important ones have already been accounted in the
value of the EDM enhancement coefficient $K$ we used in the present work.
In addition there are uncertainties due to 4)experimentally
unknown $E_{4f}-E_{5d}$ energy splitting in Gd$^{3+}$, and 5)unaccounted
exchange diagrams in Fig.6.
The most important is the energy splitting
$E_{4f}-E_{5d}$ that determines uncertainty in
the electron EDM enhancement coefficient $K$, see eq.(\ref{K}).
We estimate the accuracy of the results
(\ref{f29}), (\ref{EE}), and (\ref{ddee}) as 30-40\%.
Experimental and/or theoretical determination of the 
$E_{4f}-E_{5d}$ energy splitting in Gd$^{3+}$
would be the most important to improve the accuracy of
the calculation of the effects related to the electron EDM.

\section{conclusions}
In the present work we have considered the T,P-odd effects in solids.
The calculations are performed for  Gadolinium Gallium
Garnet, Gd$_3$Ga$_5$O$_{12}$ (GGG), and Gadolinium Iron Garnet 
Gd$_3$Fe$_5$O$_{12}$ (GdIG). We consider the effects due to the
electron electric dipole moment (EDM) and due to the Nuclear Schiff Moment 
(NSM). Both GdIG and GGG have uncompensated electron spins on Gd$^{3+}$
ions. There are two possibilities to probe the electron EDM.
The first one is to polarize magnetically the electron spins and to measure
the induced voltage across the sample. According to our calculations
at the current limitation on the electron EDM, (\ref{Tl}), the induced voltage 
across a 10cm sample is $0.54\times 10^{-9}V$, see eq. (\ref{EE}).
Another possibility is to apply an electric field to the unpolarized sample.
This leads to the spin-dependent energy shift of each Gd ion $\Delta \epsilon =
2.8\times 10^{-22} eV$ at $E=10kV/cm$, see eq. (\ref{ddee}). This can be 
measured via macroscopic magnetization of the sample.

Gd nuclei can be polarized by the electron polarization via the hyperfine
interaction or they can be polarized independently. For
100\% polarization at the current limitation on the NSM (\ref{Hg}) the 
induced voltage across a 10cm sample is $0.9\times 10^{-11}V$, see eq. 
(\ref{ES}). Application of the external electric field to the unpolarized
sample leads to the  energy shift of each Gd $\Delta \epsilon =
0.47\times 10^{-23} eV$ at $E=10kV/cm$, see eq. (\ref{ddss}). 
The shift is dependent on the nuclear spin. This can be 
measured in NMR experiments or via macroscopic electronic magnetization of 
the sample due to the hyperfine interaction.

Another possibility for searches of NSM is to use compounds without unpaired
electrons, such as  Lutetium Gallium Garnet, Lu$_3$Ga$_5$O$_{12}$
or Lutetium Oxide Lu$_2$O$_3$. In this case both the NMR and the macroscopic 
magnetization techniques can be used. A possibility to cool the nuclear spin 
subsystem down to very low temperatures \cite{NT} can give an additional 
advantage.

In our estimates of the effects induced by NSM  we
have used the current limitation (\ref{Hg}) for NSM of $^{199}$Hg
as the reference point.
However the NSM is dependent on nucleus. $^{199}$Hg is a simple
spherical nucleus while $^{155}$Gd is a deformed nucleus with
close (105keV) levels of opposite parity 
and low energy ($\sim 1MeV$)
collective $3^-$ excitation. In this situation an 
additional enhancement of NSM by an order of magnitude is possible.
This problem requires separate consideration.

There is one more mechanism to probe T,P-violation inside nucleus.
It is related to the T,P-odd nuclear electric octupole moment and it also 
requires
separate consideration. This mechanism was mentioned in Ref \cite{SFK},
but has never been considered. For $^{155}$Gd it can give a large
contribution because of the low energy collective $3^-$ excitation.

\acknowledgments
We are grateful to S. K. Lamoreaux,  L. R. Hunter, and D. Budker who
attracted our attention to the problem. We are also grateful to them as
well as to W. R. Johnson and M. G. Kozlov for stimulating discussions
and interest in the work.


\begin{references}
\bibitem{c} We found the effective principle numbers by direct comparison 
of the Relativistic Hartree-Fock wave functions at small distances with the 
corresponding semiclassical formulas for a neutral atom, 
see, e.g. Ref.\cite{KL}. This is why we do not  account for the effective 
charge of the ion $Z_i=2$ in eqs. (\ref{sp}),(\ref{spz}), and (\ref{spz1}).
One can account for $Z_i$, but then the effective
principle numbers have to be rescaled, $\nu \to \mu$: $1/\nu^3=Z_i^2/\mu^3$.
The matrix element is invariant under this procedure.
\end{references}
\end{document}